\documentclass[twocolumn,letter]{jpsj3} 

\usepackage{amsmath}

\title{Anisotropic Transport Properties of CeRu$_2$Al$_{10}$}

\author{
Hiroshi \textsc{Tanida}\thanks{E-mail address: tany@hiroshima-u.ac.jp}, 
Daiki \textsc{Tanaka}, Masafumi \textsc{Sera}, Chikako \textsc{Moriyoshi}$^1$, Yoshihiro \textsc{Kuroiwa}$^1$\\ Tomoaki \textsc{Takesaka}$^2$, Takashi \textsc{Nishioka}$^2$, Harukazu \textsc{Kato}$^2$, and Masahiro \textsc{Matsumura}$^2$
}

\inst{
Department of Quantum Matter, ADSM, Hiroshima University, Higashi-Hiroshima, Hiroshima 739-8530

$^1$Department of Physics, Hiroshima University, Higashi-Hiroshima, Hiroshima 739-8530

$^2$Graduate School of Integrated Arts and Science, Kochi University, Kochi 780-8520
}

\abst{
We studied the transport properties of CeRu$_2$Al$_{10}$ single crystals.
All the transport properties show largely anisotropic behaviors below $T_0$.
Those along the $a$- and $c$-axes show similar behaviors, but are different from those along the $b$-axis.
This suggests that the system could be viewed as a two-dimensional system.
The results of the thermal conductivity and thermoelectric power could be explained by assuming the singlet ground state below $T_0$.
However, the ground state is not simple but has some kind of structure within a spin gap.  

}

\kword{CeRu$_2$Al$_{10}$, spin gap, singlet ground state, Kondo semiconductor}

\begin{document}
\maketitle


The ternary rare-earth compound CeRu$_2$Al$_{10}$ has attracted much attention because of the mysterious transition at $T_0$=27 K.\cite{Strydom,Nishioka,Matsumura}
As for the long-range order (LRO) below $T_0$, Strydom proposed magnetic ordering.\cite{Strydom}
However, Nishioka $et~al$. pointed out that $T_0$=27 K is too high for RRu$_2$Al$_{10}$ with R=Ce because, even in GdRu$_2$Al$_{10}$, $T_{\rm N}$=16 K; they insisted that the origin of the LRO below $T_0$ is nonmagnetic and proposed charge density wave (CDW) formation.\cite{Nishioka}
Matsumura $et~al$. showed no evidence of magnetic ordering below $T_0$ from the results of $^{27}$Al-NQR measurement and proposed the occurrence of structural phase transition below $T_0$.\cite{Matsumura}
In our previous paper,\cite{Tanida} we studied the La substitution effect on CeRu$_2$Al$_{10}$ and the magnetic field effect on $T_0$ in order to clarify the nature of the LRO \cite{Tanida}, and found that $T_0$ is reduced by La substitution and disappears at $x\sim$0.45 in Ce$_x$La$_{1-x}$Ru$_2$Al$_{10}$.
The suppression of $T_0$ by a magnetic field along the easy magnetization axis was also observed.
From these results, we concluded that the origin of the LRO below $T_0$ is magnetic.
Considering the above points and a large magnetic entropy at $T_0$, a decrease in the magnetic susceptibility below $T_0$ along all the crystal axes, the existence of a gap in many physical quantities, and a positive pressure effect on $T_0$, we proposed that the LRO with a singlet ground state is formed below $T_0$.\cite{Tanida}
However, many physical properties remain to be explained from this standpoint, as mentioned in our previous paper.\cite{Tanida}
In the present study, we measured the transport properties of CeRu$_2$Al$_{10}$ single crystal, focusing on their anisotropy to clarify the unusual nature in the LRO below $T_0$.

Single crystals of Ce$_x$La$_{1-x}$Ru$_2$Al$_{10}$($x$=1 and 0.1) and CeFe$_2$Al$_{10}$ were prepared by the Al self-flux method.
Hall resistivity was measured by the conventional four probe method below $\sim$50 K.
Thermal conductivity was measured by the conventional steady state method below $\sim$ 50 K.
Thermoelectric power was measured by the conventional differential method between 1.8 and 300 K.


Figure 1 shows the temperature ($T$) dependence of the Hall resistivity $\rho_{\rm H}$ of CeRu$_2$Al$_{10}$ measured at $H$=5 T, which is applied along the $a$-, $b$-, and $c$-axes.
$\rho_{\rm H}$ shows a roughly $H$ linear dependence up to 5 T and the sign of $\rho_{\rm H}$ is positive in the entire temperature range studied here.
$\rho_{\rm H}$ shows a large anisotropy depending on the applied magnetic field direction.
$\rho_{\rm H}$ for $H\parallel b$ is the largest and that for $H\parallel a$ is the smallest.
$\rho_{\rm H}$ for $H\parallel b$ increases with decreasing temperature in a paramagnetic region and increases largely after showing a kink at $T_0$.
$\rho_{\rm H}$ below $T_0$ shows no convex $T$ dependence, which is expected when the gap is opened on the Fermi surface but shows a concave one. 
For $H\parallel c$, the $T$ dependence of $\rho_{\rm H}$ in the entire temperature range is similar to that for $H\parallel b$, although its magnitude is much smaller than that for $H\parallel b$.
The magnitude at the lowest temperature is about one third of that for $H\parallel b$.
Noted that the anomaly at $T_0$ is much less pronounced than that for $H\parallel b$ and a small hump is seen at approximately $T$$\sim$17 K.
For $H\parallel a$, $\rho_{\rm H}$ is the smallest and the $T$ dependence above $\sim$ 20 K is very different from those for $H\parallel b$ and $c$.
No anomaly is recognized at $T_0$ and a concave $T$ dependence is seen down to $\sim$17 K below $\sim$25 K; in a paramagnetic region, its $T$ dependence is very small and its magnitude is also much smaller than those for $H\parallel b$ and $c$. 
We emphasize that $\rho_{\rm H}$ below $T_0$ for $H{\parallel}b$ where the electrical current flows in the $ac$-plane shows very different behaviors from those for $H{\parallel}c$ and $a$. 
$\rho_{\rm H}$ for $H{\parallel}c$ is largest and that for $H{\parallel}a$ is smallest in the whole temperature range studied here. 
Although $\rho_{\rm H}$ for $H{\parallel}b$ and $c$ show a kink at $T_0$, that for $H{\parallel}a$ shows no anomaly. 
Although the $T$ dependence is seen for $H{\parallel}b$ and $c$ in a paramagnetic region, $\rho_{\rm H}$ for $H{\parallel}a$ is independent of temperature. 
If we estimate carrier concentration assuming a single carrier, we will obtain values of $\sim$0.1, $\sim$0.02, and $\sim$0.06/unit cell at $T$=1.5 K for $H\parallel a$, $b$, and $c$, respectively.
At $T$=35 K, they are $\sim$6.3, $\sim$1.2, and $\sim$2.0/unit cell for $H\parallel a$, $b$, and $c$, respectively.

\begin{figure}[tb]
\begin{center}
\includegraphics[width=0.55\linewidth]{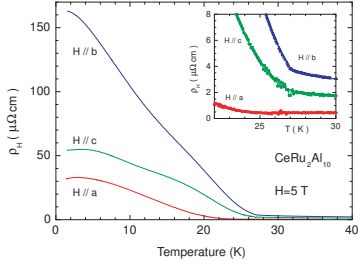}
\end{center}
\caption{
(Color Online) Temperature dependence of Hall resistivity of CeRu$_2$Al$_{10}$ measured at $H$=5 T, where data were obtained by following the configurations of (i) $H{\parallel}a$, $I{\parallel}c$, $V_{\rm H}{\parallel}b$, (ii) $H{\parallel}b$, $I{\parallel}a$, $V_{\rm H}{\parallel}c$, and (iii) $H{\parallel}c$, $I{\parallel}a$, $V_{\rm H}{\parallel}b$, respectively. Here, $V_{\rm H}$ is the Hall voltage.
}
\label{f1}
\end{figure}

Figure 2 shows the $T$ dependence of the thermal conductivity $\kappa$ along the $a$-, $b$-, and $c$-axes. 
The $T$ dependences and magnitudes of $\kappa_a$ and $\kappa_c$ are similar to each other but different from those of $\kappa_b$.
$\kappa_a$ and $\kappa_c$ show a sharp dip at $T_0$ and a clear enhancement below $T_0$. 
After showing a broad maximum at approximately $T{\sim}$24 K, $\kappa_a$ and $\kappa_c$ decrease with decreasing temperature. 
Although $\kappa_b$ also shows a dip at $T_0$, it is much less pronounced than in $\kappa_a$ and $\kappa_c$ and no enhancement below $T_0$ is seen.
$\kappa$ along the three crystal axes is roughly proportional to the temperature below $\sim$ 12 K down to 2 K.
A negative intersect is obtained if it is extrapolated down to $T$=0.
At low temperatures below 2 K, $\kappa$ is expected to show a $T^3$ dependence, which originates from the specific heat of phonon because the sound velocity and phonon mean free path are expected to be independent of temperature at low temperatures.
Since the magnitude of the electrical resistivity $\rho$ is large, the electronic contribution to $\kappa$ is small.
Even in the case with the smallest $\rho$ along the $a$-axis, the electronic contribution is estimated to be $\sim$2\% at $T$=20 K assuming the Wiedemann-Franz law.
Thus, thermal current is mainly carried by phonon.
Then, the observed anisotropy of $\kappa$ is considered to be dominated by the anisotropy of the phonon contribution.

\begin{figure}[tb]
\begin{center}
\includegraphics[width=0.55\linewidth]{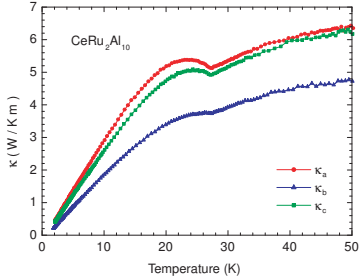}
\end{center}
\caption{
(Color Online) Temperature dependence of thermal conductivity of CeRu$_2$Al$_{10}$ along the $a$-, $b$-, and $c$-axes.
}
\label{f1}
\end{figure}

Figure 3 shows the $T$ dependence of the thermoelectric power $S$ of Ce$_{0.1}$La$_{0.9}$Ru$_2$Al$_{10}$ along the $a$-, $b$-, and $c$-axes.
$S$ is negative at high temperatures and is positive at low temperatures.
All the results show similar $T$ dependences and their magnitudes are also close to each other.
At high temperatures, $S$ shows a roughly $T$-linear dependence with a negative slope.
At $T\sim$ 10 K, a distinct positive peak is observed.
Considering the results of $\rho$ and specific heat of this sample, this positive peak is expected to originate from the Kondo effect.
A very broad hump at $\sim$150 K may result from the crystalline-electric-field (CEF) effect. \cite{Fulde,Maekawa}

\begin{figure}[tb]
\begin{center}
\includegraphics[width=0.55\linewidth]{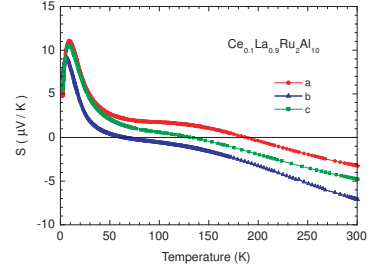}
\end{center}
\caption{
(Color Online) Temperature dependence of thermoelectric power of Ce$_{0.1}$La$_{0.9}$Ru$_2$Al$_{10}$ along the $a$-, $b$-, and $c$-axes.
}
\label{f1}
\end{figure}

Figure 4 shows the $T$ dependence of $S$ of CeRu$_2$Al$_{10}$ along the $a$-, $b$-, and $c$-axes.
All the results show a broad maximum at $T\sim$150 K, which is much more pronounced than those of Ce$_{0.1}$La$_{0.9}$Ru$_2$Al$_{10}$.
This maximum may reflect the CEF splitting, which is expected to be larger than $\sim$ 300 K.
Although a change of sign is observed in $S_c$ between $\sim$25 K and $\sim$100 K, and the magnitude of maximum differs between $S_a$, $S_c$, and $S_b$, the overall $T$ dependences are similar to each other along three crystal axes in a paramagnetic region.
Namely, a broad maximum at high temperatures of approximately ${\sim}$150 K, a decrease down to $\sim$50 K, and an increase down to $T_0$.
The increase in $S$ below 50 K down to $T_0$ seems to be associated with the large increase in $\rho$ in the same temperature region.
$S$ below $T_0$ shows a very anisotropic and anomalous $T$ dependence as is shown in the inset of Fig. 4.
The most characteristic feature is that although $S_a$ and $S_c$ show a steep increase below $T_0$, $S_b$ shows a steep decrease below $T_0$.
This again suggests that the nature in the $ac$-plane is different from that along the $b$-axis.
Below $T_0$, $S_a$ shows a pronounced maximum at $T\sim$22 K and decreases with decreasing temperature.
A pronounced shoulder is seen at $T\sim$8 K.
$S_c$ also shows a steep increase below $T_0$. 
After showing a very small concave curvature at ${\sim}$14 K, it shows a maximum at $T{\sim}$6 K. 
On the other hand, $S_b$ shows a steep decrease below $T_0$ and after exhibiting a broad minimum at $T\sim$18 K, it shows a maximum at $T\sim$ 6 K.
Thus, $S$ shows two characteristic behaviors. 
One is a steep increase in $S_a$ and $S_c$ and a steep decrease in $S_b$ below $T_0$.
Another is the existence of a broad maximum at 6$\sim$8 K along all the crystal axes.

\begin{figure}[tb]
\begin{center}
\includegraphics[width=0.55\linewidth]{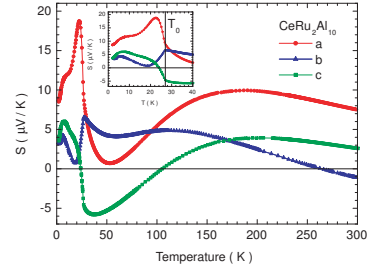}
\end{center}
\caption{
(Color Online) Temperature dependence of thermoelectric power of CeRu$_2$Al$_{10}$ along the $a$-, $b$-, and $c$-axes.
}
\label{f1}
\end{figure}


The present results of $\rho_{\rm H}$, $\kappa$, and $S$ strongly suggest that the origin of the mysterious LRO below $T_0$ exists in the $ac$-plane.
First, we consider the nature of LRO below $T_0$ from the standpoint of the crystal structures.
The powder X-ray diffraction was examined by BL02B2 in SPring-8.
The lattice constants of CeRu$_2$Al$_{10}$ are $a$=9.12233(4), $b$=10.27486(4), and $c$=9.18353(4)${\rm {\AA}}$ and those of CeFe$_2$Al$_{10}$ are $a$=9.00628(6), $b$=10.23105(7), and $c$=9.07838(6)${\rm {\AA}}$ at room temperature.
These values are obtained by Rietveld analysis.
Although the valence of the Ce ion in CeRu$_2$Al$_{10}$ is considered to be +3, the Ce ion in CeFe$_2$Al$_{10}$ is considered to be situated in a valence fluctuation regime from the magnetic susceptibility.\cite{Muro,Nishioka}
In our previous study,\cite{Nishioka} we showed that the $\rho$ of CeRu$_2$Al$_{10}$ becomes closer to that of CeFe$_2$Al$_{10}$ with increasing pressure.
This suggests that some hint on the LRO in CeRu$_2$Al$_{10}$ could be obtained from the difference of the lattice constants between these two compounds.
Here, we consider the ratio (Ru/Fe) of each lattice constant of CeRu$_2$Al$_{10}$ to that of CeFe$_2$Al$_{10}$.
The ratios are 1.01289, 1.00428, and 1.01147 for the $a$-, $b$-, and $c$-axes, respectively.
The magnitude of shrinkage in the $ac$-plane is more than twice as large as that along the $b$-axis.
In the $ac$-plane, the magnitude of shrinkage for the $a$-axis is nearly the same as that for the $c$-axis.
Considering these results, we conjecture that the LRO below $T_0$ mainly originates from the $ac$-plane.

Here, we describe the characteristics of the crystal structure in detail.
Figures 5(a) and 5(b) show the crystal structures of CeRu$_2$Al$_{10}$, which are viewed from the $c$- and $a$-axes, respectively.  
In these figures, Ru-Al 1$\sim$Al 5 are connected by lines.
By considering the present results showing that the transport properties in the $ac$-plane are different from those along the $b$-axis, the above-mentioned anisotropy of lattice constants and the fact that the Al 5 site is a specific site from the NQR measurement, we consider that a layer formed by Ce, Ru, and Al 1$\sim$Al 4 is separated by Al 5 ions along the $b$-axis.
Although Al 5 ions lie on a straight line along the $b$-axis, as shown in Fig. 5(b), when viewed from the $a$-axis, the ions form a zigzag structure along the $b$-axis when viewed from the $c$-axis, as shown in Fig. 5(a). 
In the upper and lower planes of this layer, there exist Ce and Al 1$\sim$Al 4 ions.
Ru ions are situated in a middle of this layer along the $b$-axis.
Although Ce ions in the upper and lower planes are equivalent, here, we distinguish them using different colors (black and gray) to see the physical meaning easily in the present discussion.
It should be noted that Ce and Al 1$\sim$Al 4 ions do not lie on the same plane.
Figure 5(c) shows the crystal structure viewed from the $b$-axis.
The Ce ions on the upper plane are surrounded by a large square formed by Al 2 and Al 3 ions.
A small square formed by Al 1 and Al 4 ions which are situated just above Ce ion in a lower plane, is situated just in the middle of four large squares.
Large and small squares are connected by two kinds of rhombuses formed by Al 1-Al 3 and Al 2-Al 4 ions.
Here, the two kind of squares are slightly distorted and the two kinds of rhombuses are not equivalent, which leads to a small difference between the lattice constants $a$ and $c$.
Figure 5(d) shows a schematic picture of the crystal viewed from the [10$\overline{1}$] direction.
In this figure, it is seen that a layer formed by Ce, Ru, and Al 1$\sim$Al 5 ions is constructed by a trapezoid with the staggered arrangement along the [101] direction.
Ce ions are situated in the middle of the base of a trapezoid.
Thus, the crystal structure of CeRu$_2$Al$_{10}$ could be viewed as a two-dimensional system.

\begin{figure}[tb]
\begin{center}
\includegraphics[width=0.6\linewidth]{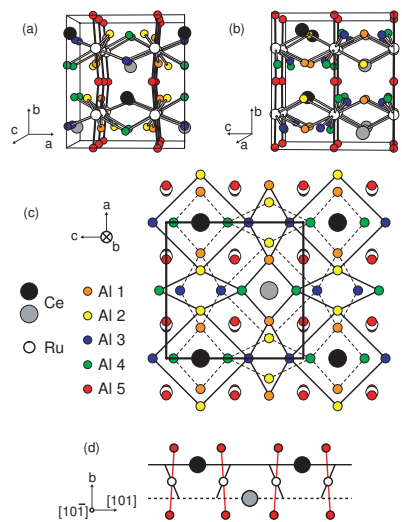}
\end{center}
\caption{
(Color Online) Crystal structure of CeRu$_2$Al$_{10}$ viewed from (a) $c$-axis, (b) $a$-axis, (c) $b$-axis, and (d) [10$\overline{1}$] direction. Although all the Ce ions are equivalent, here, we distinguish Ce ions in the upper plane (black circle) from those in the lower plane (gray circle) of the layer formed by Ce, Ru, and Al 1$\sim$ Al 4.
The upper and lower planes are drawn by a solid and dotted lines, respectively.
 
}
\label{f1}
\end{figure}


Now, we discuss the present results.
First, we discuss thermal conductivity.
As mentioned above, the contribution of the charge carrier can be neglected and thermal current is mainly carried by phonons.
Then, $\kappa$ is expressed as $C_{ph}v_sl_{ph}/3$.
Here, $C_{ph}$, $v_s$, and $l_{ph}$ are the specific heat of phonon, sound velocity, and mean free path of phonon, respectively.
Since $C_{ph}$ does not depend on the crystal axis, the anisotropy of $\kappa$ originates from that of $v_s$ or $l_{ph}$.
At high temperatures, $\kappa_b$ is the smallest.
This could be explained by considering the existence of the zigzag alignment of the Al 5 site along the $b$-axis, which is shown in Fig. 5(a).
At around $T_0$, although a clear dip at $T_0$ and a clear enhancement below $T_0$ are observed in $\kappa_a$ and $\kappa_c$, the anomalies at $T_0$ in $\kappa_b$ are much less pronounced.
These suggest that the origin of the LRO lies in the $ac$-plane, and a small anomaly along the $b$-axis seems to appear as a secondary effect.
In our previous paper,\cite{Tanida} we proposed that the enhancement of $\kappa$ below $T_0$ could be explained as follows. 
The absolute value of $\rho$ is large and so $\kappa$ is dominated by phonons. 
Below $T_0$, by forming a singlet pair between Ce ions, the vibration of Ce ions is expected to be suppressed more than when each Ce ion vibrates independently above $T_0$.  
As a result, the mean free path of phonons could be enhanced below $T_0$. 
This causes the enhancement of $\kappa$ below $T_0$. 
At present, we do not know how $v_s$ behaves at around $T_0$. 
The above should be confirmed in the future studies.


Next, we discuss the Hall effect.
Before discussing the Hall effect, we mention the results of $\rho$.
$\rho$ shows a steep increase below $T_0$ along all the crystal axes.\cite{Nishioka}
This simply suggests that the gap is opened on the Fermi surface rather isotropically.
However, such a situation could occur only when the almost complete nesting of the Fermi surface takes place.
However, this is quite difficult to expect in the present three-dimensional complex sample.
Furthermore, if such a complete isotropic nesting takes place below $T_0$, $\rho_{\rm H}$ should increase below $T_0$ in all the applied field directions.
However, no such behavior is observed.
No anomaly is seen in $\rho_{\rm H}$ for $H\parallel a$ at $T_0$.
This denies the gap opening at least for $H\parallel a$ and suggests that it is better to find other origins of the anomalous transport properties below $T_0$, except the gap opening on the Fermi surface.
$\rho_{\rm H}$ shows the clearest anomaly at $T_0$ for $H\parallel b$ and the largest magnitude.
Here, current flows in the $ac$-plane.
This suggests that the transport property in the $ac$-plane is most largely affected by the LRO at $T_0$.
In a paramagnetic region, although $\rho_{\rm H}$ shows an increase with decreasing temperature down to $T_0$ for $H\parallel a$ and $c$, that for $H\parallel b$ is temperature-independent in the temperature range studied here.
The former could be ascribed to a decrease in carrier concentration with decreasing temperature, which seems to be consistent with the $T$ dependence of $\rho$, or to the anomalous Hall effect.
However, no $T$ dependence of $\rho_{\rm H}$ for $H\parallel a$ where current flows in the $bc$-plane is observed, and its magnitude is much smaller than those for $H\parallel b$ and $c$ above $T_0$.
Thus, both below and above $T_0$, it is difficult to estimate the real carrier concentration in CeRu$_2$Al$_{10}$.
Further studies are necessary to clarify the origin of the anisotropic behavior of $\rho_{\rm H}$.


Next, we discuss thermoelectric power, which shows the most anisotropic behavior below $T_0$ among the three transport properties studied here.
$S$ is expressed as

\begin{equation}
\displaystyle S=\frac{1}{eT}\frac{\int\sigma(\varepsilon)(\varepsilon-\zeta)\frac{\partial f(\varepsilon)}{\partial \varepsilon}d\varepsilon}{\int \sigma(\varepsilon)\frac{\partial f(\varepsilon)}{\partial \varepsilon}d\varepsilon},
\end{equation}

\noindent
where, $f(\varepsilon)$ is the Fermi distribution function and $\zeta$ is the Fermi energy.
$S$ is expressed by Mott formulation as

\begin{equation}
\displaystyle S=\frac{\pi^2 k_{\rm B}^2 T}{3e}\left[\frac{1}{N(\varepsilon)v^2}\frac{\partial N(\varepsilon)v^2}{\partial \varepsilon}+\frac{1}{\tau(\varepsilon)}\frac{\partial \tau(\varepsilon)}{\partial \varepsilon}\right]_{\varepsilon=\zeta},
\end{equation}

\noindent
where $N(\varepsilon)$, $v$, and $\tau(\varepsilon)$ are the density of states, Fermi velocity, and relaxation time of conduction electrons, respectively.
In order to obtain a large magnitude of $S$, a large energy dependence of $\tau(\varepsilon)$ at around $\varepsilon=\zeta$ as a function of energy is necessary.
$\rho$ is determined by the magnitude of $\tau(\varepsilon)$ at $\varepsilon=\zeta$ itself.
In $S$, its magnitude is important but the energy dependence of $\tau(\varepsilon)$, $\partial \tau(\varepsilon)/\partial \varepsilon$ is more important.  
Namely, the inelastic scattering of charge carriers associated with the term $(\varepsilon-\zeta)$ in eq. (1) is essentially important to understand the anomalous behavior in $S$.
When the gap is opened on the Fermi surface, the first term in eq. (2) originating from $N(\varepsilon)$, which is affected by the gap opening, is also important.
However, as mentioned above, it is expected that the gap is not opened at least isotropically on the Fermi surface.
Thus, we ascribe the anomalous $T$ dependence of $S$ below $T_0$ to the energy dependence of $\tau(\varepsilon)$ at around $\varepsilon=\zeta$.
It is known that the Kondo impurity induces a large anomalous $T$ dependence in $S$ at low temperatures.
Kondo explained this unusual behavior of $S$ by considering the fourth-order perturbation $J^3 V$.\cite{Kondo}
The potential scattering term $V$ is introduced for $\tau(\varepsilon)$ to be an odd function of energy.
Peshcel and Fulde calculated $S$ for a compound with a singlet ground state by considering the third-order perturbation $J^2 V$.\cite{Fulde}
The existence of CEF splitting with a singlet ground state reduces the perturbation from the fourth order to the third order.
They showed that $S$ exhibits a Schottky-type anomaly in its $T$ dependence.
This originates from the inelastic scattering of charge carriers by a spin-flip scattering between the CEF ground state and the excited state.
Namely, in such a CEF level scheme, the higher the energy, the larger the spin-flip scattering probability because the magnetic excited state is situated on the higher energy side.
In our previous paper,\cite{Tanida} we proposed that the singlet ground state is formed below $T_0$.
The present anomalous $T$ dependence of $S$ below $T_0$ could be explained using the above mechanism proposed by Peschel and Fulde.
In the model by Peschel and Fulde, the energy splitting between two levels is temperature-independent and so a Schottky-type anomaly is obtained.
However, in the present case, the singlet-triplet splitting is induced below $T_0$, which leads to a discontinuous enhancement below $T_0$.
The difference in the sign of $dS/dT$ between the $ac$-plane and the $b$-axis just below $T_0$ could be ascribed to a difference in the sign of $V$, although its microscopic mechanism is as yet unknown.
A maximum of $S$ together with a shoulder of $\rho$ at $T\sim$8 K suggests a change in the energy dependence of $\tau(\varepsilon)$ at around this temperature.
This suggests the existence of some kind of structure within a spin gap.


To conclude, we studied the thermal transport properties in CeRu$_2$Al$_{10}$ single crystal.
All the transport properties show very anisotropic behaviors below $T_0$.
Those in the $ac$-plane are very different from those along the $b$-axis, which suggests that the system could be viewed as a two-dimensional system.
The results of $\rho_{\rm H}$ deny a gap opening on the Fermi surface at least for $H\parallel a$ and suggest the existence of other mechanisms of the anomalous transport properties except the gap opening on the Fermi surface.
The results of $\kappa$ and $S$ below $T_0$ could be explained by assuming the LRO with a singlet ground state.
However, the ground state is not simple but has a structure within a spin gap.

\end{document}